\documentclass[aps,prd,fleqn,superscriptaddress]{revtex4}
\usepackage{graphicx,xcolor,natbib,braket,float}
\usepackage{amsmath,amssymb,amsfonts}
\newcommand{\bse}{\begin{subequations}}
\newcommand{\ese}{\end{subequations}}
\newcommand{\be}{\begin{equation}}
\newcommand{\ee}{\end{equation}}
\newcommand{\bea}{\begin{eqnarray}}
\newcommand{\eea}{\end{eqnarray}}
\newcommand{\ba}{\begin{array}}
\newcommand{\ea}{\end{array}}

\usepackage[colorlinks=true, linkcolor=blue, bookmarks=true]{hyperref}

\begin{document}
\title{Analytical HTEE in Moving Plasmas}
\author{Mohammad Ali-Akbari\footnote{$\rm{m}_{-}$aliakbari@sbu.ac.ir}}
\affiliation{Department of Physics, Shahid Beheshti University, 1983969411, Tehran, Iran}

\begin{abstract}
We investigate the holographic temporal entanglement entropy of a boosted \((1+1)\)-dimensional plasma within the AdS\(_3\)/CFT\(_2\) correspondence, considering a Lorentz-boosted BTZ black hole background. We solve the coupled extremal-surface equations analytically and obtain a closed-form expression for the HTEE. Our results show that the real part of the SHTEE depends on the velocity, whereas the imaginary part is independent of the velocity and remains constant and universal. The known static BTZ result is recovered in the zero velocity limit, as expected.
\end{abstract}

\maketitle

\tableofcontents

\section{Introduction}
According to the AdS/CFT correspondence, a quantum theory of gravity in an asymptotically anti-de Sitter (AdS$_5$) spacetime is equivalent to a conformal field theory (CFT$_4$) defined on its boundary \cite{Maldacena:1997re, Witten:1998qj, Gubser:1998bc, Aharony:1999ti}. This duality establishes a precise link between bulk gravitational dynamics and boundary field theory observables, offering a powerful tool for studying strongly coupled systems. Over the past decades, holography has found broad applications in quantum field theory, condensed matter physics \cite{Hartnoll:2009ns, Tong:2016kpv}, and quantum chromodynamics \cite{Casalderrey-Solana:2011dxg}, and has proven essential in exploring non-equilibrium processes \cite{Heller:2014wfa, Grieninger:2020wsb, Shahkarami:2017fxc, Ali-Akbari:2016aqx} and entanglement-related observables such as entanglement entropy. This framework has not only deepened our understanding of strongly correlated matter but has also revealed profound connections between quantum information and the emergence of spacetime geometry, inspiring a new perspective on the fundamental nature of gravity.

Entanglement entropy is a fundamental measure of quantum correlations in quantum field theories \cite{Headrick:2019eth, Calabrese:2004eu, Casini:2009sr}. Defined as the von Neumann entropy of a reduced density matrix, it quantifies the quantum information shared between complementary subsystems. Owing to its broad applications in condensed matter physics, quantum information theory, high-energy physics and black hole physics, entanglement entropy has become an indispensable tool for exploring the structure of quantum systems. It satisfies important properties such as subadditivity and strong subadditivity and captures universal features of quantum correlations, including scaling behaviors that are sensitive to the underlying degrees of freedom and the geometry of the entangling surface. Its computation in interacting quantum field theories, however, remains notoriously challenging, motivating the development of powerful non-perturbative approaches such as AdS/CFT correspondence or more generally holographic duality.

Holographic entanglement entropy provides a geometric prescription for computing the entanglement entropy of strongly coupled quantum field theories through the AdS/CFT correspondence \cite{Ryu:2006bv}. According to the Ryu$-$Takayanagi (RT) proposal and its covariant generalization by Hubeny, Rangamani, and Takayanagi (HRT), the entanglement entropy of a boundary subsystem is determined by the area of an extremal surface anchored on the boundary of the subsystem and extending into the bulk spacetime \cite{Hubeny:2007xt}. This remarkable relation has become one of the most successful applications of holography and has provided profound insights into the interplay between quantum information and gravity. The holographic entanglement entropy has been extensively studied in various contexts, including time-dependent backgrounds, finite-temperature systems and theories with chemical potential, leading to a deeper understanding of quantum correlations in strongly interacting systems \cite{Rangamani:2016dms, Nishioka:2009un}. These developments have not only enriched our knowledge of holographic dualities but have also inspired new perspectives on the emergence of spacetime geometry from entanglement \cite{VanRaamsdonk:2010pw}.

Temporal entanglement entropy has recently emerged as a natural extension of the conventional notion of entanglement entropy to timelike separated regions \cite{Doi:2023zaf, Mollabashi:2020yie}. Unlike the standard entanglement entropy, which quantifies quantum correlations between spatial subregions on a fixed time slice, temporal entanglement entropy probes correlations between different times and offers a new perspective on the temporal structure of quantum field theories. This concept has recently been extended to the holographic framework where the holographic temporal entanglement entropy (HTEE) is computed from complex extremal surfaces in complexified bulk spacetimes \cite{Heller:2024whi}. The emergence of HTEE has opened a new direction for investigating temporal quantum correlations within the AdS/CFT correspondence, motivating further studies of its properties in various holographic backgrounds. In particular, HTEE has been explored in the context of black hole spacetimes, out-of-equilibrium systems, and finite-temperature theories, revealing novel features that are inaccessible to spatial entanglement entropy alone \cite{Heller:2025kvp, Nunez:2025gxq, Katoch:2025bnh, Prihadi:2026nua, Jena:2024tly, Bhattacharya:2022msw, Bernamonti:2026pxo, Pal:2026ysc, Kusuki:2017jxh, Katoch:2026dzs}. These developments suggest that temporal entanglement may provide a complementary diagnostic of quantum information dynamics in strongly coupled systems with potential implications for our understanding of thermalization and the structure of spacetime.

The general framework of HTEE in terms of complex extremal surfaces was established in \cite{Heller:2024whi}. Building on this idea, in this work we compute the HTEE, namely the temporal entanglement entropy measured by a stationary (lab-frame) observer in a moving plasma. The HTEE in the presence of velocity has also been studied in \cite{Kusuki:2017jxh}. In that work, the authors consider a boosted interval in a three-dimensional asymptotically AdS$_3$ black hole, namely the BTZ black hole. This is precisely the opposite of the setup considered here: in \cite{Kusuki:2017jxh} the interval is boosted while the black hole is static, whereas in our case the time interval is static and the black hole (plasma) is boosted. Although the final result for the HTEE is expected to be the same in both setups, \cite{Kusuki:2017jxh} does not provide an analytical result for the HTEE that can be compared with ours; the authors only plot the HTEE as a function of the spacelike/timelike interval. Fortunately, we are able to solve the relevant equations of motion analytically and obtain a closed-form result for the HTEE. Therefore, our result is more instructive in this respect and reveals different properties of the HTEE, as we will see.

In this work, in order to establish a baseline for comparing the results in different limits, in section 2 we review and obtain the HTEE in the zero- and finite-temperature regimes of two-dimensional conformal field theory. In the next section, we first boost the BTZ black hole and then compute the HTEE analytically in the presence of velocity, step by step. In all cases, our time interval is taken to be at rest in the lab frame, while the plasma moves past it. In the appendices, the relevant integrals are explained and consistency checks on our result are illustrated.

While this paper was being finalized, a related paper appeared on arXiv \cite{Bernamonti:2026pxo}, presenting further developments on HTEE. It provides a boundary CFT formulation and studies temporal entanglement in a boosted conical-defect geometry. Although there are some similarities, our setup and calculation are different, and our final result is parametrized in terms of different physical quantities.

\section{The HTEE in Static Backgrounds}
As discussed, we would like to compute the HTEE in a $1+1$ dimensional moving plasma. Before doing so, in order to explain our strategy, we first consider the zero- and non-zero-temperature cases to reproduce the standard analytical HTEE available in the literature. This provides a useful consistency check of our method before extending it to the moving plasma. Therefore, we start with a three-dimensional black hole which is asymptotically $\mathrm{AdS}_3$ given by
\begin{equation}
ds^2 = \frac{1}{z^2} \left( -f(z)dt^2 + \frac{dz^2}{f(z)} + dx^2 \right),
\label{eq:metric}
\end{equation}
where $z$ denotes the radial direction and the boundary is located at $z\rightarrow 0$. The coordinates $t$ and $x$ denote the time and spatial directions on which the dual field theory lives. The blackening factor is
\begin{equation}
f(z) = 1 - \frac{z^2}{z_h^2},
\label{eq:blackening}
\end{equation}
where $z_h$ denotes the horizon of the black hole. The temperature of the black hole, which is identified with the temperature of the dual field theory, is given by $T = \frac{1}{2\pi z_h}$.

The boundary region whose HTEE we would like to compute is chosen to be
\begin{equation}
A:\{t\in [-\frac{\Delta t}{2},\frac{\Delta t}{2}],\ x=0\},
\label{eq:region}
\end{equation}
where $\Delta t$ is the time interval. Owing to the translational symmetry along the spatial direction, it is sufficient to describe the bulk extremal surface $\Gamma_{\Delta t}$ by the embedding function $t(z)$, which, in the present case, describes a geodesic in the bulk geometry. We now need to extremize the corresponding area (length) functional
\begin{equation}
L = \frac{1}{z}\sqrt{-f\dot{t}^2 + \frac{1}{f}},
\label{eq:lagrangian}
\end{equation}
where $\dot{}=\frac{d}{dz}$. Since $t(z)$ does not appear explicitly in the above Lagrangian, there exists a conserved quantity, which is obtained from
\begin{equation}
E = \frac{\partial L}{\partial \dot{t}}.
\label{eq:conserved_E}
\end{equation}
Solving this equation for $\dot{t}$ leads to
\begin{equation}
\dot{t}^2 = \frac{z^2E^2}{f^2(f + z^2E^2)}.
\label{eq:tdot_squared}
\end{equation}
The turning point $z_*$ of the extremal surface corresponds to the deepest radial point reached by the geodesic where the embedding function satisfies $\dot{t} \to \infty$. This condition yields
\begin{equation}
f(z_{*}) + z_{*}^{2}E^{2} = 0.
\label{eq:turning_static}
\end{equation}
The endpoints of the bulk geodesic are anchored on the asymptotic boundary at
\begin{equation}
t(\epsilon) = \pm \frac{\Delta t}{2},
\label{eq:boundary_endpoints}
\end{equation}and therefore
\begin{equation}
\frac{\Delta t}{2} = \int_{\epsilon}^{z_{*}} dz \, \dot{t},
\label{eq:delta_t_integral}
\end{equation}
where $\epsilon$ is the UV cutoff that regulates the location of the asymptotic boundary. \eqref{eq:delta_t_integral} establishes the relation between the integration constant $E$ and the boundary time interval $\Delta t$.

Having determined the profile of the extremal surface, the remaining task is to evaluate its regularized length, from which the HTEE can be computed according to the holographic prescription. The regularized length of the extremal geodesic is then obtained by evaluating the length functional on the solution $t(z)$. In the following, we evaluate the above integrals analytically and demonstrate that the resulting expression exactly reproduces the known results in the literature. 

\subsection{Zero temperature case}

In this subsection, we consider the zero-temperature case, corresponding to $f(z) = 1$, or equivalently the limit $z_{h} \to \infty$. In this limit, the geometry reduces to pure AdS$_3$, allowing the relevant quantities to be evaluated analytically. This provides an important consistency check of our prescription, since the corresponding HTEE is already known in the literature.

\eqref{eq:tdot_squared} is then simply reduced to
\begin{equation}
\dot{t} = \frac{zE}{\sqrt{1 + z^{2}E^{2}}}.
\label{eq:tdot_zero}
\end{equation}
Using \eqref{eq:turning_static} and \eqref{eq:delta_t_integral}, after straightforward calculations, one finds
\begin{align}
z_{*}E &= i, \label{eq:zero_zt}\\
\Delta t &= \frac{2}{E}\sqrt{1 + \epsilon^{2}E^{2}}. \label{eq:zero_dt}
\end{align}
\eqref{eq:zero_zt} implies that the turning point is complex and, consequently, the extremal surface extends into the complexified bulk geometry. \eqref{eq:zero_dt} establishes the relation between the integration constant $E$ and the boundary time interval $\Delta t$. As will be seen below, this complex turning point is essential for reproducing the correct HTEE.

We now evaluate the regularized length by substituting \eqref{eq:tdot_zero} into \eqref{eq:lagrangian}, obtaining
\begin{align}
L_{\mathrm{reg}} &= \int_{\epsilon}^{z_{*}}\frac{dz}{z}\frac{1}{\sqrt{1 + z^{2}E^{2}}}, \label{eq:length_zero_1}\\
&= \ln (z_{*}E) - \ln \left[\frac{E}{1 + \frac{E\Delta t}{2}}\right], \label{eq:length_zero_2}\\
&= \frac{i\pi}{2} +\ln \left(\frac{\Delta t}{\epsilon}\right) + \mathcal{O}(\epsilon^{2}), \label{eq:length_zero_3}
\end{align}
where we have assumed that $\Delta t \gg \epsilon$. The logarithmic divergence is regulated by the UV cutoff $\epsilon$, while the finite part contains a non-vanishing imaginary contribution arising from the complex turning point.
Substituting the regularized length into the holographic formula \cite{Heller:2024whi}
\begin{equation}
S = \frac{\mathrm{Area}(\Gamma_{\Delta t})}{4G},
\label{eq:htee_formula}
\end{equation}
we obtain
\begin{equation}
S = \frac{2L_{reg}}{4G} = \frac{c}{3}\ln \left(\frac{\Delta t}{\epsilon}\right) + i\frac{\pi c}{6},
\label{eq:htee_zero}
\end{equation}
where $c = \frac{3}{2G}$ and $G$ is Newton constant. This is precisely the well-known result obtained in the literature; see, for example \cite{Heller:2024whi}. 

We would like to emphasize that \eqref{eq:zero_zt} plays a crucial role in obtaining the above result, as can be seen from \eqref{eq:length_zero_2}. Without imposing this condition, the turning point becomes real, and although a real solution still exists, it does not reproduce the imaginary contribution to the entropy. Therefore, the complex extremal surface is essential for correctly recovering the HTEE. Furthermore, since $\Delta t > 0$, \eqref{eq:zero_dt} implies that the integration constant $E$ must be positive in order to produce a positive boundary time interval.

\subsection{Finite temperature case}

We now perform the corresponding calculations for the black hole background. In this case, it is straightforward to find
\begin{equation}
\dot{t} = \frac{zE}{f^2\sqrt{f + z^2E^2}},
\label{eq:tdot_finite}
\end{equation}
which is slightly more complicated than the zero-temperature case due to the presence of the blackening factor. To determine the turning point, we set the denominator of the above expression to zero. Using \eqref{eq:turning_static} and \eqref{eq:delta_t_integral}, after straightforward calculations, one finds
\begin{align}
z_{*} &= \frac{i}{\sqrt{E^{2} - \frac{1}{z_{h}^{2}}}}, \label{eq:finite_zt}\\
\Delta t &= z_{h}\ln \left[\frac{Ez_{h} + 1}{Ez_{h} - 1}\right]. \label{eq:finite_dt}
\end{align}
Similar to the zero-temperature case, the turning point is complex whenever $Ez_h > 1$. This condition not only guarantees the appearance of an imaginary contribution to the HTEE but also ensures that the argument of the logarithm in \eqref{eq:finite_dt} remains positive. Hence, the existence of a complex turning point is once again directly linked to the emergence of the imaginary part of the entropy.

Combining the equations in \eqref{eq:finite_zt} and \eqref{eq:finite_dt}, one obtains
\begin{equation}
z_{*} = iz_{h}\sinh \left(\frac{\Delta t}{2z_{h}}\right),
\label{eq:finite_zt_dt}
\end{equation}
which expresses the turning point directly in terms of the boundary time interval. We are now ready to evaluate the regularized length of the extremal surface. Using \eqref{eq:lagrangian}, \eqref{eq:tdot_finite} and \eqref{eq:finite_zt_dt}, one obtains
\begin{equation}
L_{reg} = \frac{i\pi}{2} +\ln \left[\frac{2z_{h}}{\epsilon}\sinh \left(\frac{\Delta t}{2z_{h}}\right)\right] + \mathcal{O}(\epsilon^{2}),
\label{eq:length_finite}
\end{equation}
where we have assumed $\Delta t \gg \epsilon$. As in the zero-temperature case, the imaginary contribution originates from the complex turning point while the logarithmic divergence is regulated by the UV cutoff $\epsilon$.
Substituting this result into \eqref{eq:htee_formula}, we obtain
\begin{equation}
S = \frac{2L_{reg}}{4G} = \frac{c}{3}\ln \left[\frac{2z_{h}}{\epsilon}\sinh \left(\frac{\Delta t}{2z_{h}}\right)\right] + i\frac{\pi c}{6}.
\label{eq:htee_finite}
\end{equation}
This expression exactly reproduces the finite-temperature result for the HTEE. In the limit $z_{h} \to \infty$, it smoothly reduces to the zero-temperature result obtained in the previous subsection.

\section{The HTEE in a Moving Plasma}

Our aim in this section is to compute the HTEE in the presence of a finite velocity. Therefore, we apply a Lorentz boost in the $x - t$ plane according to
\begin{align}
t &\rightarrow \gamma (t - vx), \label{eq:boost_t}\\
x &\rightarrow \gamma (x - vt), \label{eq:boost_x}
\end{align}
where $\gamma = \frac{1}{\sqrt{1 - v^2}}$ is the Lorentz boost factor and $v$ denotes the velocity of the plasma. This transformation allows us to describe the gravitational dual of a uniformly moving thermal state in the boundary field theory.
The metric after applying the boost takes the form
\begin{equation}
ds^{2} = \frac{1}{z^{2}}\left(\frac{dz^{2}}{f} + g_{tt}dt^{2} + 2g_{tx}dt dx + g_{xx}dx^{2}\right),
\label{eq:boosted_metric}
\end{equation}
where
\begin{align}
g_{tt} &= \gamma^2 (v^2 - f), \label{eq:gtt}\\
g_{xt} &= \gamma^2 v(f - 1), \label{eq:gtx}\\
g_{xx} &= \gamma^2 (1 - f v^2). \label{eq:gxx}
\end{align}
Notice that the boost generates an off-diagonal component $g_{tx}$, reflecting the fact that the background is no longer static in these coordinates. Consequently, the time and spatial directions become coupled, making the analysis of the extremal surface more involved than in the static case.
The boundary region whose HTEE we would like to compute in the presence of a nonzero velocity is chosen to be the same as in \eqref{eq:region}. A few important points are in order.

\begin{itemize}
\item We emphasize that the HTEE is defined with the subregion given by \eqref{eq:region}, i.e., a pure timelike interval at fixed $x = 0$ in the lab-frame, even though the plasma is moving. Physically, the HTEE is interpreted as the temporal entanglement entropy measured by a stationary detector at rest in the lab-frame, which records correlations in a moving medium. See table \ref{tab:observers_vertical}, which summarizes the two distinct observers.

\begin{table}[htbp]
\centering
\caption{Two distinct observers and their corresponding HTEEs.}
\begin{tabular}{|l|c|c|}
\hline
\textbf{Property} & \textbf{Observer 1: Static}  & \textbf{Observer 3: Stationary} \\
\hline
Quantity & Standard HTEE & Boosted HTEE \\
\hline
Velocity & $v = 0$ & $v \neq 0$ \\
\hline
Metric & Static BTZ & Boosted BTZ \\
\hline
Subregion & $t \in [-\Delta t/2, \Delta t/2]$ &  $t \in [-\Delta t/2, \Delta t/2]$  \\
\hline
Type of interval & Timelike in lab-frame & Timelike in lab-frame \\
\hline
Entropy & \eqref{eq:htee_finite} &  \eqref{eq:htee_final} \\
\hline
Imaginary part & $+\dfrac{i\pi c}{6}$ & $+\dfrac{i\pi c}{6}$  \\
\hline
\end{tabular}
\label{tab:observers_vertical}
\end{table}

\item Due to the non-vanishing off-diagonal component of the boosted metric ($g_{tx} \neq 0$), the extremal geodesic acquires a nontrivial dependence on both the temporal and spatial directions. Consequently, unlike the static case where the extremal surface is described by a single embedding function $t(z)$, the bulk extremal surface in the boosted background must be parametrized by two independent embedding functions, $t(z)$ and $x(z)$. As a result, the extremization problem becomes considerably more involved, since the length functional now depends on two functions. Nevertheless, the boosted geometry retains translational invariance along both the $t$ and $x$ directions, giving rise to two conserved quantities that significantly simplify the analysis. In the following, we derive the corresponding equations of motion and use these conserved quantities to determine the extremal geodesic and the associated HTEE.

\item As a final consistency check, we note that the HTEE reduces to the static HTEE in the zero velocity limit. This is expected, since in the limit $v \to 0$, the off-diagonal component of the metric vanishes ($g_{tx} \to 0$), and the boosted background reduces to the static BTZ geometry. Consequently, the extremal surface becomes purely temporal, and our result smoothly reproduces the known static HTEE. Throughout this manuscript, we will repeatedly use this limit as a benchmark to verify the validity of our computations.

\end{itemize} 

The corresponding length functional that must be extremized is
\begin{equation}
L = \frac{\sqrt{Q}}{z},
\label{eq:L_boosted}
\end{equation}
where
\begin{equation}
Q = \frac{1}{f} + \gamma^2 (v^2 - f)\dot{t}^2 + 2\gamma^2 v(f - 1)\dot{t}\dot{x} + \gamma^2 (1 - f v^2)\dot{x}^2.
\label{eq:Q_def}
\end{equation}
Since neither $t(z)$ nor $x(z)$ appears explicitly in the Lagrangian \eqref{eq:L_boosted}, there exist two conserved quantities associated with translations along the time and spatial directions. These constants of motion are given by
\begin{align}
E &= -\frac{\partial L}{\partial \dot{t}} = \frac{\gamma^2}{z\sqrt{Q}}\left((v^2 - f)\dot{t} + v(f - 1)\dot{x}\right), \label{eq:E_moving}\\
P &= -\frac{\partial L}{\partial \dot{x}} = \frac{\gamma^2}{z\sqrt{Q}}\left((1 - f v^2)\dot{x} + v(f - 1)\dot{t}\right). \label{eq:P_moving}
\end{align}
The existence of two conserved quantities follows from the translational symmetries generated by the Killing vectors $\partial_t$ and $\partial_x$.
Using the above equations, one can express $\dot{t}$ and $\dot{x}$ in terms of the conserved quantities as
\begin{align}
\dot{t} &= \frac{\gamma^{2}z\sqrt{Q}}{f}\left((1-fv^{2})E - v(f-1)P\right), \label{eq:tdot_moving}\\
\dot{x} &= \frac{\gamma^{2}z\sqrt{Q}}{f}\left(-v(f-1)E - (v^{2}-f)P\right). \label{eq:xdot_moving}
\end{align}
At this stage, one encounters a technical difficulty. The above equations are still coupled through the quantity $Q$ which depends explicitly on both $\dot{t}$ and $\dot{x}$. Consequently, \eqref{eq:tdot_moving} and \eqref{eq:xdot_moving} cannot be solved directly and must first be simplified.
Substituting \eqref{eq:tdot_moving} and \eqref{eq:xdot_moving} into \eqref{eq:Q_def}, one obtains
\begin{equation}
Q = \frac{1}{f + \gamma^2 z^2 s(z)},
\label{eq:Q_simplified}
\end{equation}
where
\begin{equation}
s(z) = (1 - f v^2)E^2 + (v^2 - f)P^2 - 2v(f - 1)EP.
\label{eq:s_z}
\end{equation}
This expression completely eliminates the dependence of $Q$ on the derivatives of the embedding functions, allowing the equations of motion to be written solely in terms of the conserved quantities.

It is now evident that either of the equations in \eqref{eq:tdot_moving} or \eqref{eq:xdot_moving} can be used to determine the turning point and both lead to the same condition. Therefore, the turning point is determined by
\begin{equation}
f(z_{*}) + \gamma^{2}z_{*}^{2}s(z_{*}) = 0.
\label{eq:turning_moving}
\end{equation}
As expected, this equation naturally reduces to the turning-point condition \eqref{eq:turning_static} in the static limit $v\rightarrow 0$, for which $P\rightarrow 0$ and $\gamma \rightarrow 1$. Furthermore, in the same limit, \eqref{eq:tdot_moving} consistently reduces to \eqref{eq:tdot_squared}. These observations demonstrate that the boosted formalism smoothly reproduces the known static results.

Solving \eqref{eq:turning_moving} for the turning point, we obtain
\begin{equation}
z_{*}^{2} = \frac{1}{\frac{2\gamma^{2}}{z_{h}^{2}}(P + Ev)^{2}}\left[-(E^{2} - P^{2} - \frac{1}{z_{h}^{2}})\pm \sqrt{(E^{2} - P^{2} - \frac{1}{z_{h}^{2}})^{2} - \frac{4\gamma^{2}}{z_{h}^{2}}(P + Ev)^{2}}\right],
\label{eq:zstar_squared}
\end{equation}
which clearly admits two distinct solutions. The second root does not correspond to a turning point of the extremal surface but is nevertheless essential for factorizing the quartic polynomial appearing in the radial integral. To determine the physically relevant branch, we examine the zero-velocity limit and require that the result reproduces the thermal turning point given in \eqref{eq:finite_zt}. A straightforward calculation shows that the solution corresponding to the plus sign satisfies this requirement and should therefore be selected.

In both the zero- and finite-temperature cases discussed in the previous sections, the turning point is purely imaginary. To maintain the same structure in the presence of a finite velocity, we rewrite the solution by factoring out a minus sign from the numerator. The turning point can then be expressed as
\begin{equation}
z_{*} = \frac{i z_{h}}{\sqrt{2}\gamma(P + Ev)}\left[(E^{2} - P^{2} - \frac{1}{z_{h}^{2}}) - \sqrt{(E^{2} - P^{2} - \frac{1}{z_{h}^{2}})^{2} - \frac{4\gamma^{2}}{z_{h}^{2}}(P + Ev)^{2}}\right]^{\frac{1}{2}},
\label{eq:zstar_imag0}
\end{equation}
where it is evident that the expression inside the first square root is always positive. Consequently, the reality of the quantity inside the second square root guarantees that the turning point remains purely imaginary.
Therefore, we restrict our analysis to the regime in which the turning point is purely imaginary. This requires the condition
\begin{equation}
(E^{2} - P^{2} - \frac{1}{z_{h}^{2}}) > \frac{2\gamma}{z_{h}}(P + Ev).
\label{eq:condition_imag}
\end{equation}
Under this assumption, the quantity inside the brackets is positive and therefore the turning point is purely imaginary. Furthermore, in the static limit $v\rightarrow 0$, this condition naturally reduces to the familiar constraint
\begin{equation}
Ez_{h} > 1,
\label{eq:ezh_condition}
\end{equation}
which follows directly from \eqref{eq:finite_zt} and \eqref{eq:finite_dt}. 

It should be noted that replacing \(i\) with \(-i\) in \eqref{eq:zstar_imag0} yields another acceptable solution for \(z_*\). As extensively discussed in \cite{Heller:2024whi}, this ambiguity arises when using complexified geometries to compute the HTEE. For the minus-sign branch, the real part of the entropy remains unchanged, while the imaginary part flips sign. In this manuscript, we adopt the plus sign for the remainder of our computation, although the opposite branch remains a valid solution.

Having determined the turning point, we next evaluate the boundary conditions. Substituting \eqref{eq:Q_simplified} into \eqref{eq:tdot_moving} and \eqref{eq:xdot_moving} and integrating from the turning point to the asymptotic boundary, we obtain
\begin{align}
\frac{\Delta t}{2} &= \gamma^2\left[(1 - v^2)E I_1 + \frac{v(P + Ev)}{z_h^2} I_2\right], \label{eq:bc1}\\
0 &= \gamma^2\left[\frac{P + Ev}{z_h^2} I_2 + (v^2 - 1)P I_1\right], \label{eq:bc2}
\end{align}
where
\begin{align}
I_1 &= \int_{\epsilon}^{z_{*}} \frac{z dz}{f \sqrt{D}}, \label{eq:I1_def}\\
I_2 &= \int_{\epsilon}^{z_{*}} \frac{z^3 dz}{f \sqrt{D}}, \label{eq:I2_def}
\end{align}
and
\begin{equation}
D = f + (E^2 - P^2)z^2 + \frac{\gamma^2}{z_h^2} (P + Ev)^2 z^4.
\label{eq:D_def}
\end{equation}
Although two different integrals appear in \eqref{eq:bc1} and \eqref{eq:bc2}, the two equations combine in a remarkably simple manner. In particular, the integral $I_2$ can be eliminated algebraically, leaving only the integral $I_1$ which is considerably simpler to evaluate. As a result, the boundary time interval can be expressed solely in terms of $I_1$ as
\begin{equation}
\Delta t = 2(E + P v) I_1.
\label{eq:delta_t_I1}
\end{equation}
This simplification significantly reduces the complexity of the calculation and enables the remaining analysis to be carried out analytically. The evaluation of the integral $I_1$ is presented in Appendix \ref{appa}, where it is computed in closed form. Substituting the resulting expression into \eqref{eq:delta_t_I1}, we obtain the following relation between the boundary time interval and the integration constants
\begin{equation}
\Delta t = \frac{z_h}{\gamma} \ln \left[\frac{1 + X}{1 - X}\right],
\label{eq:delta_t_ln}
\end{equation}
where
\begin{equation}
X = \frac{\sqrt{z_h^2 - z_r^2}}{\sqrt{z_h^2 - z_*^2}} \frac{z_*}{z_r}.
\label{eq:X_def}
\end{equation}
The above equation can be readily inverted to give
\begin{equation}
X = \tanh \left(\frac{\gamma \Delta t}{2 z_h}\right).
\label{eq:X_tanh}
\end{equation}
This relation will prove particularly useful in simplifying the final expression for the regularized length since it allows all dependence on the integration constants to be expressed in terms of the physical boundary time interval.

Interestingly, as demonstrated in Appendix B, \eqref{eq:delta_t_ln} correctly reduces to \eqref{eq:finite_zt_dt} in the zero velocity limit. This provides an additional consistency check of our analysis and confirms that the boosted solution smoothly interpolates to the static black hole background. \eqref{eq:delta_t_ln} therefore establishes the desired relation between the integration constants and the boundary time interval $\Delta t$. Together with the turning-point condition, it provides all the ingredients required to evaluate the regularized length of the extremal surface and, consequently, the HTEE.

We now proceed to compute the regularized length of the extremal geodesic,
\begin{equation}
L_{reg} = \int_{\epsilon}^{z_{*}} \frac{\sqrt{Q} dz}{z}.
\label{eq:L_reg_integral}
\end{equation}
Substituting the corresponding expressions for the conserved quantities and the extremal trajectory into the above integral, and after straightforward algebraic simplifications, we obtain
\begin{equation}
L_{reg} = \frac{i\pi}{2} + \ln \left[\frac{2 z_h |z_r|}{\epsilon \sqrt{z_h^2 + |z_r|^2}} \sinh \left(\frac{\gamma \Delta t}{2 z_h}\right)\right],
\label{eq:L_reg_moving}
\end{equation}
where we have introduced
\begin{equation}
z_{r} = \frac{i z_{h}}{\sqrt{2}\gamma(P + Ev)}
\left[
(E^{2} - P^{2} - \frac{1}{z_{h}^{2}}) + 
\sqrt{(E^{2} - P^{2} - \frac{1}{z_{h}^{2}})^{2} - \frac{4\gamma^{2}}{z_{h}^{2}}(P + Ev)^{2}}
\right]^{\frac{1}{2}},
\label{eq:zstar_imag}
\end{equation}

There are two limits we would like to check before proceeding further:

\begin{itemize}
\item From Appendix~\ref{appb}, it is clear that in the zero velocity limit the above expression reduces precisely to \eqref{eq:length_finite}.

\item In the zero-temperature limit, notice that $|z_r|\sim \frac{z_h}{\gamma v}$ and $\sinh x \sim x$. Moreover, using the fact that translational invariance is restored in this limit, we can, without loss of generality, set $P=0$. With these considerations, we reproduce \eqref{eq:length_zero_3}.
\end{itemize}

Finally, substituting the regularized length into the holographic prescription for the temporal entanglement entropy \eqref{eq:htee_formula}, we arrive at
\begin{equation}
S_{\Delta t}^{\rm HTEE} = \frac{2L_{reg}}{4G} = \frac{c}{3}\ln \left[\frac{2|z_r|}{\epsilon\sqrt{1 +(2\pi T)^2 |z_r|^{2}}}\sinh \left(\pi\gamma T \Delta t\right)\right] + i\frac{\pi c}{6}.
\label{eq:htee_final}
\end{equation}
\eqref{eq:htee_final} constitutes the central result of this work. It explicitly captures the dependence of the HTEE on the velocity of the dual plasma. Moreover, the persistence of the universal imaginary contribution demonstrates that the complex nature of the extremal surface remains intact in the presence of the boost, while the real part acquires a nontrivial velocity dependence.

The parameter $|z_r|$ is not an independent physical quantity but is determined implicitly by the equations defining the extremal geodesic. In principle, $|z_r|$ can be eliminated in favor of the boundary time interval $\Delta t$, temperature and the constants. However, this requires solving a quartic algebraic equation, which considerably obscures the final expression without providing additional physical insight. For this reason, we retain $|z_r|$ as an implicit parameter throughout our analysis. \eqref{eq:htee_final} therefore provides the most compact closed-form expression for the HTEE.

The computation of the HTEE in the present work follows the same holographic prescription as \cite{Heller:2024whi}, namely the evaluation of the regularized length of complex extremal geodesics in a complexified bulk geometry. In both approaches, the complexification of the spacetime and the corresponding complex extremal surfaces are essential for generating the imaginary contribution to the HTEE. The main difference lies in the formulation of the variational problem. While \cite{Heller:2024whi} begins with the reparameterization-invariant geodesic action and derives the equations of motion before fixing a parameterization, we formulate the problem directly in terms of the gauge-fixed length functional by taking the radial coordinate $z$ as the parameter along the extremal curve. 

\begin{figure}[H]
  \centering
  \includegraphics[width=0.55\textwidth]{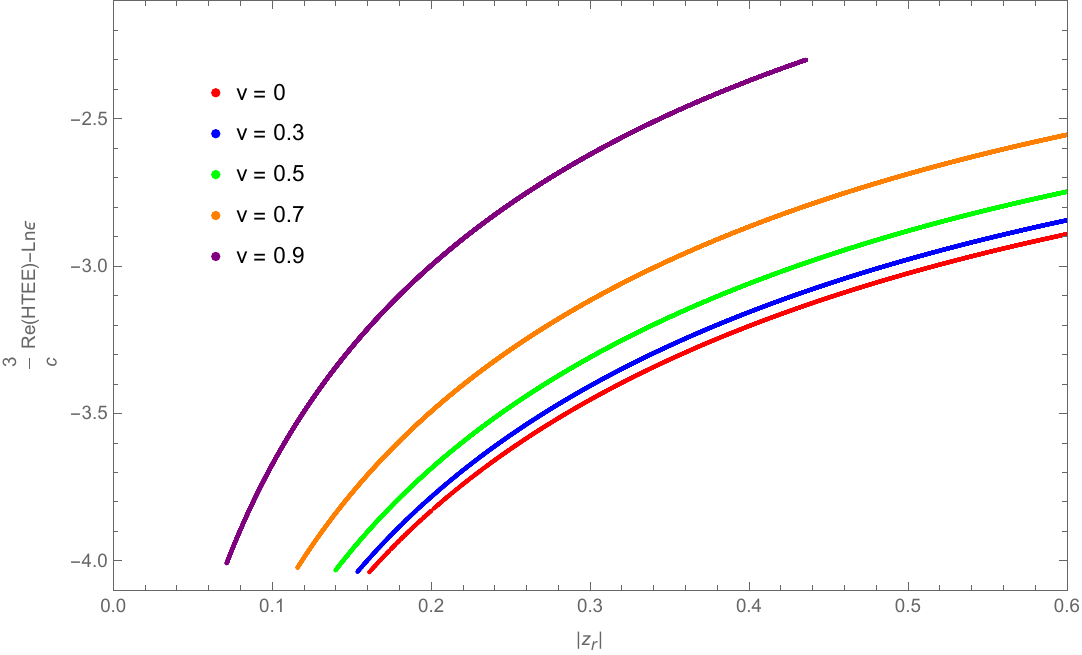}
  \caption{HTEE as a function of $|z_r|$ for different boost velocities, with $\Delta t = 0.1$ and $T = 0.177$ fixed. Only constant pairs ($E$, $P$) satisfying the inequality \eqref{eq:condition_imag} and the boundary condition \eqref{eq:X_tanh} are shown.}
  \label{fig}
\end{figure}

In figure \ref{fig}, we plot the HTEE as a function of \(|z_r|\) for various values of the boost velocity, while keeping the time interval \(\Delta t\) and the temperature \(T\) fixed. The figure reveals a clear monotonic increase of the HTEE with velocity for any given value of \(|z_r|\). This behavior can be understood as a direct consequence of the plasma flow: the Lorentz boost enhances the real part of the HTEE, thereby leading to the observed increase.
It is important to emphasize that the parameters \(E\) and \(P\) appearing in the construction are not arbitrary. They must simultaneously satisfy the inequality condition \eqref{eq:condition_imag}, which ensures that the turning point remains purely imaginary, and the boundary condition \eqref{eq:X_tanh}, which fixes the physical time interval \(\Delta t\). These two conditions severely restrict the allowed parameter space, ensuring that only the physically admissible branches of the complex geodesic are included in the figure. Consequently, the data points shown correspond to a well-defined subset of extremal surfaces consistent with the complexification prescription.
Finally, we note that while the real part of the HTEE exhibits this nontrivial velocity dependence, the universal imaginary contribution \(i\pi c/6\) remains completely unaffected by both the boost parameter and the value of \(|z_r|\).

Analysing \eqref{eq:delta_t_ln}, one finds
\begin{equation}
\Delta t \in \mathbb{R} \quad \Longrightarrow \quad X \in \mathbb{R}.
\end{equation}
Using \eqref{eq:X_def}, it is straightforward to see that \(X\) is real if and only if both ratios
\be
\frac{z_*}{z_r}
\quad\text{and}\quad
\frac{\sqrt{z_h^2 - z_r^2}}{\sqrt{z_h^2 - z_*^2}}
\ee
are real. This condition holds in precisely two distinct cases: either \(z_*\) and \(z_r\) are both real, or they are both purely imaginary. In either case, the product of the two ratios is real, which ensures that \(\Delta t\) remains real. Consequently, any complex turning point with both nonzero real and imaginary parts is inadmissible in our computation. This simple argument shows that the reality of the physical time interval \(\Delta t\) enforces a strict branch choice for the turning points, restricting the allowed configurations to either the purely imaginary branch, which yields the HTEE, or the purely real branch. The latter is analogous to the spacelike holographic entanglement entropy, which we denote by HEE\(^t\). The superscript \(t\) indicates that, unlike the standard holographic entanglement entropy, which is computed for a spacelike interval, the interval considered here is timelike. In other words, besides a complex extremal surface, we obtain a real solution corresponding to the same time interval. We defer a more detailed discussion to a forthcoming paper.

\appendix
\section{Evaluation of the Integral $I_1$}\label{appa}
From \eqref{eq:I1_def} and \eqref{eq:D_def}, the integral \(I_{1}\) can be written as
\begin{equation}
I_{1} = \int_{z_{*}}^{\epsilon}\frac{z dz}{(1 - \frac{z^{2}}{z_{h}^{2}})\sqrt{1 + Az^{2} + Bz^{4}}},
\label{eq:I1_integral}
\end{equation}
where
\begin{equation}
A = E^{2} - P^{2} - \frac{1}{z_{h}^{2}},\qquad B = \frac{\gamma^{2}}{z_{h}^{2}} (P + Ev)^{2}.
\label{eq:A_B}
\end{equation}
To evaluate this integral analytically, it is convenient to factorize the quartic polynomial appearing under the square root. Using the relation
\begin{equation}
1 + Az^{2} + Bz^{4} = B(z^{2} - z_{*}^{2})(z^{2} - z_{r}^{2}),
\label{eq:factorization}
\end{equation}
the denominator can be expressed as the product of two quadratic factors. Here, \(z_{r}\) denotes the second root of the quartic polynomial which is determined from
\begin{align}
-\frac{A}{B} &= z_{*}^{2} + z_{r}^{2}, \label{eq:zr_relation1}\\
\frac{1}{B} &= z_{r}^{2}z_{*}^{2}. \label{eq:zr_relation2}
\end{align}
Since the turning point \(z_{*}\) is purely imaginary, the above relations imply that the second root \(z_{r}\) is also purely imaginary. Consequently, the integral takes the form
\begin{equation}
I_{1} = \frac{z_{h}^{2}}{\sqrt{B}}\int_{z_{*}}^{\epsilon}\frac{z dz}{(z_{h}^{2} - z^{2})\sqrt{(z^{2} - z_{*}^{2})(z^{2} - z_{r}^{2})}}.
\label{eq:I1_simplified}
\end{equation}
In this form, the integral can be evaluated by performing a sequence of elementary changes of variables. Specifically, we introduce the transformations
\begin{align}
z^{2} &= u, \label{eq:trans1}\\
u &= z_{*}^{2}\sin^{2}\theta, \label{eq:trans2}\\
\cos \theta &= -y, \label{eq:trans3}\\
y &= \frac{\sqrt{b}}{z_{*}}\sinh w, \label{eq:trans4}\\
\tanh w &= q, \label{eq:trans5}
\end{align}
where
\begin{equation}
a = z_{h}^{2} - z_{*}^{2},\qquad b = z_{r}^{2} - z_{*}^{2}.
\label{eq:a_b}
\end{equation}
After carrying out the above sequence of variable transformations, the original integral is reduced to the much simpler form
\begin{equation}
I_{1} = \frac{z_{h}^{2}}{\sqrt{B}}\int \frac{dq}{a + (b - a)q^{2}},
\label{eq:I1_reduced}
\end{equation}
which can be evaluated analytically using a standard integral formula. To identify the appropriate form of the solution, it is necessary to determine the sign of the coefficient multiplying \(q^{2}\). Since both the turning point \(z_{*}\) and the second root \(z_{r}\) are purely imaginary, one finds that
\begin{equation}
a > 0,\qquad b - a < 0.
\label{eq:a_b_signs}
\end{equation}
It is therefore convenient to rewrite the denominator as
\begin{equation}
a + (b - a)q^{2} = a - |c|q^{2},
\label{eq:denom_abs}
\end{equation}
where \(|c| = a - b > 0\). The integral then takes the form
\begin{equation}
I_{1} = \frac{z_{h}^{2}}{\sqrt{B}}\int \frac{dq}{a - |c|q^{2}},
\label{eq:I1_standard}
\end{equation}
which is of the standard type
\begin{align}
\int \frac{dk}{n_1 - n_2 k^2} &= \frac{1}{\sqrt{n_1 n_2}}\tanh^{-1}\left(\sqrt{\frac{n_2}{n_1}} k\right) + \mathrm{constant}, \label{eq:standard_int1}\\
&= \frac{1}{2\sqrt{n_1 n_2}}\ln \left(\frac{1 + \sqrt{\frac{n_2}{n_1}} k}{1 - \sqrt{\frac{n_2}{n_1}} k}\right) + \mathrm{constant}. \label{eq:standard_int2}
\end{align}
Applying this result directly, we obtain
\begin{equation}
I_{1} = \frac{z_{h}^{2}}{2\sqrt{B a |c|}}\ln \left(\frac{1 + \sqrt{\frac{|c|}{a}} q}{1 - \sqrt{\frac{|c|}{a}} q}\right).
\label{eq:I1_result}
\end{equation}
Thus, the integral \(I_{1}\) is obtained in closed form. It is worth emphasizing that the appearance of the inverse hyperbolic tangent is a direct consequence of the negative sign in front of the quadratic term in the denominator. Had this coefficient been positive, the corresponding result would instead involve the inverse tangent function. This distinction will play an important role in determining the analytic structure of the HTEE.

\section{Zero Velocity Limit of the Time Interval}
\label{appb}
We begin by taking the zero velocity limit of the general expression for the time interval given in \eqref{eq:delta_t_I1}. In this limit, the momentum vanishes, $P \to 0$,  and the Lorentz factor tends to unity, $\gamma \to 1$. Consequently, the boundary condition \eqref{eq:delta_t_I1} reduces to
\begin{equation}
\Delta t = 2E I_1.
\label{eq:B_dt}
\end{equation}
To evaluate \(I_{1}\) in this static regime, we first determine the asymptotic behavior of its constituent parameters. From the definitions in Appendix \ref{appa}, we find
\begin{align}
A &= E^2 - \frac{1}{z_h^2}, \label{eq:B_A}\\
z_*^2 &= -\frac{1}{A}, \label{eq:B_zr}\\
a &= z_h^2 - z_*^2 = \frac{E^2 z_h^2}{A}, \label{eq:B_a}\\
B &= \frac{E^2 v^2}{z_h^2}, \label{eq:B_B}\\
z_r^2 &= -\frac{A}{B} = -\frac{A z_h^2}{E^2 v^2}, \label{eq:B_zr2}\\
|c| &= a - b \simeq \frac{A z_h^2}{E^2 v^2}. \label{eq:B_c}
\end{align}
Substituting these limiting expressions into the general formula for \(I_{1}\) given in \eqref{eq:I1_result}, the prefactor simplifies dramatically:
\begin{equation}
\sqrt{B a |c|} = \sqrt{\frac{E^2 v^2}{z_h^2} \cdot \frac{E^2 z_h^2}{A} \cdot \frac{A z_h^2}{E^2 v^2}} = E z_h.
\label{eq:prefactor}
\end{equation}
Meanwhile, the argument of the logarithm evaluates to
\begin{equation}
\frac{1 + \sqrt{\frac{|c|}{a}} q}{1 - \sqrt{\frac{|c|}{a}} q} = \frac{1 + \frac{1}{E z_h}}{1 - \frac{1}{E z_h}} = \frac{E z_h + 1}{E z_h - 1}.
\label{eq:log_arg}
\end{equation}
Collecting these results, the integral \(I_{1}\) becomes
\begin{equation}
I_{1} = \frac{z_{h}^{2}}{2(E z_{h})}\ln \left(\frac{E z_{h} + 1}{E z_{h} - 1}\right) = \frac{z_{h}}{2E}\ln \left(\frac{E z_{h} + 1}{E z_{h} - 1}\right).
\label{eq:I1_static}
\end{equation}
Finally, substituting this expression back into \eqref{eq:B_dt} yields the well-known static time interval
\begin{equation}
\Delta t = 2E\cdot \frac{z_h}{2E}\ln \left(\frac{E z_h + 1}{E z_h - 1}\right) = z_h\ln \left(\frac{E z_h + 1}{E z_h - 1}\right),
\label{eq:B_result}
\end{equation}
which precisely matches \eqref{eq:finite_dt} of the main text. This demonstrates that the boosted formalism smoothly reduces to the established static result, providing a robust consistency check of our analysis.

\end{document}